\documentstyle[12pt]{article} 
\newcommand{\be}{ \begin{equation}}
\newcommand{\ee}{\end{equation}} 
\begin{document} 
\def\theequation{\arabic{section}.\arabic{equation}} 
\begin{titlepage} 
\title{Negative energy and stability in scalar-tensor gravity} 
\author{Valerio Faraoni\\ \\ 
{\small \it Physics Department, University of Northern 
British Columbia}\\ 
{\small \it 3333 University Way, Prince George, B.C., Canada V2N~4Z9}\\ 
{\small \it  email~~vfaraoni@unbc.ca} }
\date{} \maketitle 
\thispagestyle{empty} 
\vspace*{1truecm} 
\begin{abstract} 
Linearized gravitational waves in Brans-Dicke and scalar-tensor theories carry 
negative energy. A gauge-invariant analysis shows that the background Minkowski
 space is 
stable at the classical level with respect to linear 
scalar and tensor inhomogeneous perturbations.
\end{abstract} 
\vspace*{1truecm} 
\begin{center} PACS: 04.20.Cv, 04.30.-w
\end{center} 
\end{titlepage}

\def\theequation{\arabic{section}.\arabic{equation}}

\section{Introduction}
\setcounter{equation}{0}

It has recently been shown \cite{BarceloVisser00} that all the standard energy
 conditions 
can  easily be violated  at the classical level in the theory of a scalar field 
coupled 
nonminimally with the spacetime curvature. Although there are some ambiguities 
in the definition of energy density and effective pressure \cite{BellucciFaraoni}, 
the possibility of violating the energy conditions is undeniable.
Even allowing for the possibility of exotica such as traversable wormholes made 
possible by the violation of the energy conditions, one would like to preserve
the non-negativity of the energy density, at least at the classical level. However,
even this last requirement may be violated in nonminimally coupled scalar field theory.
 Even worse, the problem of negative energy fluxes is not 
unique to nonminimally coupled theory \cite{footnote1} --- it also occurs when 
linearized gravitational waves are considered in Brans-Dicke or 
more general scalar-tensor 
theories \cite{VF04,FaraoniGunzigIJTP,FaraoniGunzigAA,BellucciFaraoniBabusciPLA}.  
To summarize the issue, consider linearized gravitational waves 
in Brans-Dicke theory, which is described by the action
\be 
S^{(BD)}=\int d^4x \, \sqrt{-g} \left[ \phi \, R -\, 
\frac{\omega}{\phi} \, g^{ab} \nabla_a \phi \, \nabla_b \phi -V( \phi) \right]
+ S^{(matter)} \;.
\ee
In the linearized version of the theory with $V(\phi)=0$ the 
metric and scalar field are given by
\be  \label{metric}
g_{ab}=\eta_{ab}+h_{ab} \;, \;\;\;\;\;\;\;\;\;\;\; \phi=\phi_0+\varphi \;,
\ee
where $\eta_{ab}$ is the Minkowski metric, $\phi_0$ is a constant 
and O$( h_{ab} ) =\mbox{O}( 
\varphi /\phi_0 ) = \mbox{O}( \epsilon )$, with $\epsilon $ a smallness 
parameter. The corresponding linearized field 
equations in a  region outside sources are
\be  \label{1}
R_{ab}= \frac{\partial _a \partial_b \varphi}{\phi_0} +\mbox{O}( \epsilon^2 ) \;,
\ee
\be \label{2}
\Box \varphi \equiv \eta^{ab} \partial_a\partial_b \varphi =0 +\mbox{O}( \epsilon^2 ) \;.
\ee
It is natural to interpret the right hand side of eq.~(\ref{1}) as an effective 
energy-momentum tensor $T_{ab}[ \varphi ] $ of the Brans-Dicke field. More generally, the 
interpretation of the right hand side of the vacuum Brans-Dicke field equations as an 
effective energy-momentum tensor is widespread in the literature and may ultimately 
be questionable 
\cite{VF04,SantiagoSilbergleit,Torres}. In fact, writing the vacuum Brans-Dicke field 
equations as 
\be
G_{ab}= \frac{1}{\phi} T_{ab}[ \phi ] = \frac{1}{\phi} \left[ \frac{\omega}{\phi} \left( 
\nabla_a \phi \nabla_b \phi -\, \frac{1}{2} g_{ab} \nabla^c \phi \nabla_c \phi \right) 
+\nabla_a \nabla_b \phi -g_{ab} \Box \phi -\, \frac{V}{2} g_{ab} \right] 
\ee
means forcing upon them an interpretation as effective Einstein equations, 
which they are not 
-- they are field equations of a theory 
different from general relativity. However, there 
is little doubt that this interpretation is appropriate in the linearized 
approximation. Let us  
consider a monochromatic component of the Fourier decomposition of 
$\varphi\left( t, \vec{x} 
\right) $ 
\be  \label{mono}
\varphi_{\vec{k}} = \varphi_0 \, \cos \left( k_c x^c \right) \;.
\ee
The effective energy density associated with the monochromatic wave (\ref{mono}) by an 
observer with four-velocity $u^a$ is 
\be
\rho = T_{ab} \, [ \varphi ] \, u^a\, u^b= -\left( k_a \, u^a \right)^2 \frac{ 
\varphi_{\vec{k}} }{ \phi_0} \;.
\ee
Due to the non-canonical form of $T_{ab}[ \varphi] $ --- 
linear in the second derivatives instead 
of quadratic in the first derivatives --- $\rho$ is not 
positive definite but oscillates with 
the frequency of $\varphi_{\vec{k}}$. This has the disturbing
 consequence that scalar-tensor 
waves emitted by a binary massive stellar system such as, e.g., $\mu$-Sco, carry a negative energy 
flux over macroscopic times (of order $3\cdot 10^5$~s for $\mu$-Sco). The contribution of the 
tensor modes $h_{ab}$ is described by the Isaacson effective tensor and 
is absent to order O($\epsilon$). 

The argument showing the presence of negative energy presented in the context of 
 Brans-Dicke theory is generalized to arbitrary scalar-tensor 
theories of gravity with gravitational sector described by the action
\be  \label{STaction}
S^{(ST)}= \int d^4 x \, \sqrt{-g} \left[ \frac{ f ( \phi )}{2} \, R -\frac{\omega( \phi )}{2}\, 
g^{ab}  \nabla_a \, \phi \, 
\nabla_b \,\phi -V( \phi) \right] \;, 
\ee
by expanding the coupling functions $ f ( \phi) $ and $\omega( \phi) $  around their 
present day values $f_0$ and $\omega_0$. 
From a conceptual point of view it could be objected 
that the consideration of Minkowski 
space is inconsistent with the original motivation of Brans-Dicke theory 
(distant matter in the universe determines the effective gravitational coupling 
$G_{eff}=\phi^{-1}$ here and now, according to Mach's principle). However, 
Minkowski space 
is a perfectly legitimate solution of the Brans-Dicke field equations from the mathematical 
point of view. Moreover, current interest in scalar-tensor gravity is not motivated by Mach's 
principle but rather by the fact that scalar-tensor theories mimic properties of stringy  
physics \cite{VF04,FujiiMaeda}. For example, the low energy limit of the bosonic string theory is  
Brans-Dicke gravity with parameter $\omega=-1$ ~\cite{string}.

The presence of negative 
energy 
fluxes is seen by certain authors as a reason to abandon the usual Jordan frame version of the 
theory and consider instead its Einstein frame counterpart with fixed 
units of time, length, and mass (see 
\cite{VF04,review,MagnanoSokolowski} for reviews). Little matters that the two 
conformally related formulations are equivalent if one allows the Einstein frame units of 
mass, time, and length to scale appropriately, as done in Dicke's original paper \cite{Dicke} 
introducing the Einstein frame version of Brans-Dicke theory. Most of the 
current literature  considers instead a version of Brans-Dicke theory in the Einstein frame 
with {\em fixed} units of mass, length, and time. 
The result is a new theory physically 
inequivalent to the original Jordan frame; in this new theory  
the scalar has canonical (positive 
definite) kinetic energy. In this paper
 we do not seek escape to the Einstein frame  but we 
work in the Jordan frame. Physicists shy away from 
negative energy because it is usually associated 
with instability and runaway solutions and intuitively this should also be the case
for classical Brans-Dicke theory and its scalar-tensor generalizations.  
It comes therefore as  a surprise that, as we show in the 
following,
 the Minkowski space taken as the background metric in 
eqs.~(\ref{metric})--(\ref{2}) is 
stable against inhomogeneous perturbations (scalar 
and tensor) to first order and 
that there are no runaway solutions to this order.

\section{Stability of Minkowski spacetime}
\setcounter{equation}{0}

Inhomogeneous perturbations are gauge-dependent and a stability 
analysis using gauge-independent quantities is mandatory. These are
 conveniently obtained by 
regarding  Minkowski space as a trivial case of a 
Friedmann-Lemaitre-Robertson-Walker 
(hereafter  ``FLRW'') universe, for which there is a vast literature 
on gauge-independent 
perturbations. Recently \cite{desitter} we have carried out a stability 
analysis of de Sitter 
spaces in  generalized gravity theories described  by the action
\be 
S = \int d^4 x \, \sqrt{-g} \left[ \frac{1}{2} \, \psi ( \phi, R )  -\frac{1}{2} \, \omega( 
\phi ) g^{ab} \nabla_a \, \phi \,  \nabla_b \,\phi -V( \phi) \right] \;,
\ee
which contains the scalar-tensor action (\ref{STaction}) as a special case. We employed the 
covariant and gauge-invariant formalism developed by Bardeen 
\cite{Bardeen}, Ellis, Bruni, 
Hwang and  Vishniac \cite{EllisBruni,HwangVishniac} in general relativity, in a 
version extended to encompass generalized gravity by Hwang 
and Hwang and Noh \cite{Hwang}.  
The gauge-invariant variables used are the Bardeen \cite{Bardeen} potentials $\Phi_H $ and 
$\Phi_A$ and the  Ellis-Bruni \cite{EllisBruni} variable $\Delta \phi$ defined by
\be \label{phiH}
\Phi_H = H_L +\frac{H_T}{3} +\frac{ \dot{a} }{k} \left( B-\frac{a}{k} \, \dot{H}_T \right) \;, 
\ee

\be \label{phiA}
\Phi_A = A  +\frac{ \dot{a} }{k} \left( B-\frac{a}{k} \, \dot{H}_T \right)
+\frac{a}{k} \left[ \dot{B} -\frac{1}{k} \left( a \dot{H}_T \right)\dot{}  \right] \;, 
\ee

\be \label{deltaphi}
\Delta \phi = \delta \phi  +\frac{a}{k} \, \dot{\phi}  \left( B-\frac{a}{k} \, \dot{H}_T 
\right) 
\;.
\ee
where $a$ is the scale factor of the background FLRW metric with line element
\be
ds^2=-dt^2 +a^2(t) \left( dx^2 +dy^2 +dz^2 \right)
\ee
and $A,B, H_L$ and $H_T$ are the 
metric perturbations defined by
\begin{eqnarray} 
g_{00} & = & -a^2 \left( 1+2AY \right) \;, \label{19} \\
&& \nonumber \\
g_{0i} & = & -a^2 \, B \, Y_i  \;, \label{20} \\
&& \nonumber \\
g_{ij} & =& a^2 \left[ h_{ij}\left(  1+2H_L \right) +2H_T \, Y_{ij}  \right] \;. \label{21}
\end{eqnarray}
Here $h_{ij} $ is the three-dimensional metric of the FLRW background and the operator $ 
\bar{\nabla_i} $ 
is the covariant derivative associated with $h_{ij}$. The scalar harmonics 
$Y$ 
are the eigenfunctions of the eigenvalue problem
\be \label{22}
\bar{\nabla_i}
\bar{\nabla^i} \, Y =-k^2 \, Y \;,
\ee
while the vector and tensor 
harmonics $Y_i$ and $Y_{ij}$ are defined by
\be \label{23}
Y_i= -\frac{1}{k} \, \bar{\nabla_i} Y \;, \;\;\;\;\;\;\;\;
Y_{ij}= \frac{1}{k^2} \, \bar{\nabla_i}\bar{\nabla_j} Y +\frac{1}{3} \, Y \, h_{ij} \;.
\ee
 The general equations for inhomogeneous perturbations \cite{Hwang} 
simplify considerably in a 
Minkowski background, reducing to 
\be  \label{onebox}
\Delta \ddot{\phi}  + \left[  k^2
 - \frac{ \left( \psi_0'' - 2V_0'' \right) }{ 2 \omega_0  
\left(  1 + 
\frac{ 3 \psi_{\phi R}^2}{2\omega_0  \psi_R^{(0)} } \right)} \right]
 \, \Delta \phi =0 \;,
\ee

\be \label{twoboxes}
\ddot{H}_T  + k^2  \, H_T=0 \;,
\ee

\be  \label{31}
\dot{\Phi}_H = - \frac{1}{2} 
 \frac{  \Delta \dot{\psi}_R }{\psi_R} \;,
\ee

\be   \label{32}
\Phi_A + \Phi_H =  - \frac{\Delta \psi_R}{ \psi_R } \; ,
\ee

where 
\begin{eqnarray} 
\psi_0 \equiv  \left. \psi \left( \phi, R \right)\right|_{ \left( 
\phi_0, R_0 \right)} 
\;,  
\;\;\;\;\;\; 
{\psi_0}'  \equiv   \left. \frac{ \partial \psi }{ \partial \phi } 
\right|_{\left( \phi_0, R_0 \right)} \;,  
\;\;\;\;\;\; 
{ \psi_0}''  \equiv  \left. \frac{\partial^2 \psi}{\partial \phi^2} 
\right|_{\left( \phi_0, 
R_0 \right)} \;,\\
&& \nonumber \\
\psi_R \equiv  \frac{\partial \psi}{\partial R} 
\;, \;\;\;\;\;\;
\psi_R^{(0)} \equiv  \left. \frac{\partial \psi}{\partial R} 
\right|_{\left( \phi_0, R_0 \right)}
 \;, \;\;\;\;\;\;
\psi_{\phi R} \equiv  \left. \frac{\partial^2 \psi}{\partial \phi\partial R} 
\right|_{\left( \phi_0, R_0 \right)} \;,
\end{eqnarray}
and $\Delta \psi_R $ is defined analogously to eq.~(\ref{deltaphi}).
 An overdot denotes differentiation with respect to the proper 
time $t$ of the FLRW background. In the cosmological 
analysis $ 
\left( H_0, \phi_0 \right)$ is  the de Sitter fixed point of which one wants to study the 
stability and $ R_0 = 12 H_0^2 $. Minkowski space corresponds to the trivial case 
 $H_0 =0$ and eqs.~(\ref{31}) 
and ({\ref{32}) yield
\be
\Phi_H=\Phi_A=-\, \frac{\Delta \psi_R}{2\psi_R^{(0)}} 
=-\frac{\psi_{\phi R} }{ 2 \psi_R^{(0)}} \, \Delta \phi \;.
\ee
Hence, we are interested in eqs.~(\ref{onebox}) and (\ref{twoboxes}) 
for the scalar and 
tensor perturbations $ \Delta \phi $ and 
$H_T$. It is obvious that the solutions of 
eq.~(\ref{twoboxes}) are 
oscillating for any real value of $k$, and hence Minkowski space is 
always stable with respect 
to 
tensor perturbations. Let us turn now to eq.~(\ref{onebox}): stability is 
equivalent to
\be
k^2+ \frac{ \left( 2V_0'' - \psi_0'' \right) }{ 2 \omega_0  
\left(  1 + \frac{ 3 \psi_{\phi R}^2}{2\omega_0 \psi_R^{(0)}}\right) } 
   \geq 0 \;.
\ee
In scalar-tensor gravity $ \psi( \phi, R ) =f( \phi) R $ and  
\be
\psi_{\phi R}= \left. \frac{df}{d\phi}\right|_{\phi_0} \;, \;\;\;\;\;\;\;
\psi_{ R}^{(0)}=  f (\phi_0) \;, \;\;\;\;\;\;\;
{\psi_0}'' = \left. \frac{d^2f}{d\phi^2} \, R \right|_{\left( \phi_0, R_0 \right) }=0 \; .
\ee
In the case of non self-interacting ($V \equiv 0$) linearized  Brans-Dicke scalar 
$\varphi$ of eq.~(\ref{metric}), Minkowski space is always stable with respect 
to linear inhomogeneous perturbations. The same conclusion holds for massive scalar waves ($V_0'' 
=m^2 >0$) if $\omega_0 >0$ and $f( \phi_0) >0$, which is the usual range of parameters in 
Brans-Dicke theory. Runaway potentials with $V_0'' < 0$ 
give unstable scalar perturbations if the wavelength is larger than a critical 
value --- the usual phenomenon present in particle dynamics with 
runaway potentials.

\section{Discussion and conclusions}
\setcounter{equation}{0}

Scalar-tensor theories are plagued by negative energies. Although it 
is not always clear how to unambigously identify energy densities 
\cite{BellucciFaraoni}, it is clear that negative energies are present 
in these theories. The situation of linearized Brans-Dicke theory
considered in Sec.~1 is free of such ambiguities --- the effective energy 
density of scalar waves can be clearly identified and 
 is not positive definite. One would therefore expect the background Minkowski
space to be unstable and to be destroyed by small perturbations. However 
this is not the case: the negative energy associated with linearized, massless, 
scalar-tensor gravitational waves does not cause  
instability or runaway 
solutions at the classical level --- Minkowski space is 
stable with respect to inhomogeneous scalar and tensor 
perturbations at the linear order. The reason for stability can be traced to the fact
that  the energy density of each 
individual mode can be negative but is  bounded from below 
once the wave frequency and 
amplitude are fixed. By contrast, one expects an instability when the 
energy of the system keeps decreasing and decreasing {\em ad infinitum}.
Hence the issue is not whether the effective energy of the scalar is negative, but
whether it has a lower bound or not. It is not trivial to answer this question 
in general because the scalar
 and the tensor fields are explicitly coupled in the full field equations, while they 
decouple to linear order. As a consequence one expects an exchange of energy and momentum 
between the scalar and the tensor field, and there is no fully satisfactory solution 
to the problem of  energy localization for the gravitational field even in general 
relativity.

Instabilities may appear at the second or higher  order 
or when the theory is quantized. 
However, it is well known that also the full equations of Brans-Dicke 
cosmology admit stable solutions. Their stability has been 
checked only with respect to homogeneous 
perturbations in several studies of the phase space, but the analysis goes
well beyond the linear order \cite{phasespace}.

Regarding quantization, the covariant perturbation scheme only works
in the Einstein conformal frame with fixed units and is not possible in 
the Jordan frame (\cite{VF04} and references therein). Thus it would
appear that the conformal transformation to the Einstein frame is a panacea for 
scalar-tensor gravity. However, the new theory in the Einstein frame with
fixed units of mass, length, and time is physically inequivalent to the original one 
in the Jordan frame \cite{footnote2}. Moreover,
 the original theory in the Jordan frame  
and its scalar-tensor generalizations are still accepted  as viable theories and 
are the subiect of a vast literature.
Nevertheless, in order for scalar-tensor theories to be fully satisfactory 
 at least at the classical
level it would be desirable to have a better  
understanding of the issues of negative  
energies and stability beyond linear order and homogeneous and isotropic 
 cosmology.

\section*{Acknowledgments}

The author acknowledges a referee for useful suggestions. 
This work was supported by the Natural Sciences and Engineering Research Council of 
Canada ({\em NSERC}).

   
\end{document}